\documentclass[useAMS,usenatbib]{mn2e}
\usepackage{graphicx}

\newcommand{\msun}{\mbox{${\rm M}_{\odot}$}}

\title[GCE in SAMs II]
{Galactic chemical evolution in hierarchical formation models - II. The Intracluster 
Medium.} 
\author[M. Arrigoni et al.]
{Mat\'{i}as Arrigoni$^1$\thanks{email: arrigoni@astro.rug.nl}, Scott C. 
  Trager$^1$ and Rachel S. Somerville$^{2,3}$ \\ 
$^1$Kapteyn Astronomical Institute, University of Groningen, Postbus 800, 
  NL-9700 AV Groningen, The Netherlands\\ 
$^2$Space Telescope Science Institute, 3700 San Martin Drive, Baltimore, 
  MD 21218, USA\\ 
$^3$Department of Physics and Astronomy, Johns Hopkins University, Baltimore, 
  MD 21218, USA}
\begin{document}

\date{Accepted . Received .}

\pagerange{\pageref{firstpage}--\pageref{lastpage}} \pubyear{2010}

\maketitle

\label{firstpage}

\begin{abstract}

We use the cosmological semi-analytic model (SAM) for galaxy formation
presented in Paper I to study the metallicities and abundance ratios
of the intracluster medium (ICM) within the hierarchical structure
formation paradigm.  By requiring a slightly flat IMF ($x=1.15$) and a
two-population delay-time-distribution (DTD) for SN Ia explosions we
found previously that this model is able to reproduce the abundance
ratios and supernova rates of early-type galaxies in the local
Universe.  Predictions for elemental abundances in the ICM pose a
further test of the model.  We find that with the fiducial model from
Paper I the overall metal content of the ICM is too low, although the
abundance ratios are in good agreement with the data. However, we find
that allowing a fraction of the metal-enriched material ejected by
stars to be deposited directly into the hot ICM, instead of being
deposited into the cold ISM, appears to be a plausible and
physically-motivated solution.

\end{abstract}


\section{Introduction}

The vast majority of baryons in the Universe reside not in stars but
in the hot and diffuse gas in clusters of galaxies---the intracluster
medium (ICM)---which provides the fuel for star formation in galaxies
\citep[e.g.][]{Lin03,Vikhlinin06}.  The metal content in this gas
suggests that it cannot be entirely of primordial origin and that a
substantial fraction must have been processed by the cluster galaxies
and then expelled back into the ICM via supernovae explosions and
galactic winds.  This interplay between the ICM and galaxies regulates
the star formation and enrichment histories \citep{renzini97} of the
Universe.  Measurement of elemental abundances in clusters can
therefore set constraints on the various feedback processes that shape
galaxy formation, as well as the relative importance of different
types of supernovae and the history of star formation. Any successful
model of galaxy formation must account not only for the observational
properties of the galaxy population but also those of the ICM.

The intergalactic medium within groups and clusters is hot and dense
enough to be observed at X-ray wavelengths. At these wavelengths Fe is
the most easily observable element. The X-ray satellites launched
since the mid-90's (ASCA, BeppoSax, Chandra, XMM-Newton) have allowed
precise measurements of many other elements such as O, Mg, Si, S, Ar,
Ca and Ni in large samples of nearby clusters
\citep{fukazawa98,peterson03,degrandi04,tamura04,dePlaa07}. As with
early-type galaxies, most of the chemical modelling of the ICM has
been done within the monolithic collapse scenario \citep{MG95, GM97};
only a handful of studies have been carried out within the
hierarchical assembly paradigm \citep{deLucia04,Nagashima05a}. These
models, however, have their own limitations. \citet{deLucia04} assume
the instantaneous recycling approximation and trace only the total
metallicity and enrichment by type II supernovae (SNe
II). \citet{Nagashima05a}, on the other hand, fully couple galactic
chemical evolution models to a SAM and successfully reproduce the
abundances of various elements in the ICM, but the same model predicts
incorrect trends of stellar abundance ratios in the early-type
galaxies within those clusters \citep{Nagashima05b}.
 
In this work, we continue our study of galactic chemical evolution
hierarchical assembly models of galaxy formation within a $\Lambda$CDM
cosmology \citep[][hereafter Paper I]{paperI}, by studying the
metallicity and abundance ratios of the hot intracluster gas. The
outline of the paper is as follows. In Section 2 we briefly describe
the semi-analytic model and the new ingredients. In Section 3 we
present our predictions and compare them with observations. In Section
4 we summarise our findings and present our conclusions.

\section{The semi-analytic model}

The backbone of our model is the SAM described by \citet[hereafter
  S08]{s08}, which tracks the hierarchical clustering of dark matter
haloes, radiative cooling of gas, star formation, SN feedback, AGN
feedback (in two distinct modes, quasars and radio jets), galaxy
mergers and the starbursts triggered by them, the evolution of stellar
populations, and the effects of dust obscuration. In Paper I, we
described our extension of this model to include detailed metal
enrichment by type Ia and type II supernovae and long-lived stars.  We
refer the reader to the two aforementioned studies for a detailed
description of the models. 

The SAM has been successful in reproducing a variety of observations
in the local Universe and at high redshift, for example, the
luminosity and stellar mass function of galaxies, the
colour--magnitude relation, galaxy star formation rates as a function
of their stellar masses, the relative numbers of early and late-type
galaxies, the gas fractions and size distributions of spiral galaxies,
and the global star formation history 
\citep[S08;][]{fontanot:09,kimm:09,hop_rss_09}.
With the addition of detailed chemical evolution modeling in Paper I,
the model is able to match the mass--metallicity relation for galaxies
and the trend of [$\alpha$/Fe] with stellar mass, as well as the
supernova rates as a function of specific star formation rate
(SSFR). To achieve this agreement, it was necessary to adopt a
Chabrier IMF \citep{Chabrier03} with a slightly flatter slope above 1
\msun ($x=1.15$ instead of $x=1.3$), a relatively low fraction of
binaries that yield a SN Ia event ($0.03$ in the $M=3$--$16\,\msun$
range), and a bimodal delay-time-distribution (DTD) with a prompt peak
and a later plateau for type Ia supernovae (SNe Ia) explosions, as
proposed by \citet{MDP06}. We will henceforth refer to the combined
GCE plus SAM as the GCE-SAM.

Here we introduce two changes relative to the GCE-SAM described in
Paper I. First, we have chosen a different set of yields for SNe
Ia. Motivated by the excessive amount of Ni present in the galaxies
and ISM in Paper I, we have switched from the yields of \citet[model
  W7]{Nomoto97} to those of \citet[model WDD3]{iwamoto99} as the
latter produces only half the Ni while the other elements remain
approximately the same. The main difference between these SN models is
the scenario for the explosion. The W7 model describes a slow
deflagration of the stellar core, while the WDD3 model is calculated
using a delayed detonation. The delayed detonation is also the
currently favoured SNIa explosion scenario \citep[see,
  e.g.,][]{dePlaa07}. The yields for SN II and AGB stars remain the
same: \citet[hereafter WW95]{WW95} and \citet{KL07}, respectively.

The other change concerns the immediate fate of the metal-rich gas
ejected by the stars. In the ``standard'' SAM of S08 (as in many SAMs,
e.g. de Lucia et al. 2004), these metals were deposited directly in
the ISM (cold gas phase associated with the individual galaxy) where
it is mixed instantaneously. However, this was a somewhat arbitrary
choice. Perhaps a more physical scenario is that the ejecta from
massive stars and supernovae is highly enriched, and it is this same
material that escapes the galaxy and pollutes the ICM. This picture is
supported by observations that indicate that galactic winds are
ubiquitously metal-enriched 
\citep{martin05,veilleux_winds05,grimes09,weiner09}, as well as by 
hydrodynamic simulations of galactic outflows 
\citep{maclow-ferrara,madau_outflow01,scannapieco_feed08}. We obtain
good results when we assume that 80\% of the new metals are deposited
directly in the hot halo gas ($f_{\rm hot\, enrich} = 0.8$). It is also
interesting to note that \citet{Li09a} find that if 95\% of the newly
produced metals are ejected directly into the hot phase for galaxies
with a DM halo mass of $5\times 10^{10}\msun$ or less, their
semi-analytic model produces a good match for the mass function and
metallicities of the Local Group dwarf satellite population. However,
our results indicate that such a mass threshold is not necessary for
reproducing the metal abundances in the ICM. 

In this paper, we adopt a flat $\Lambda$CDM cosmology with
$\Omega_{0}=0.28$, $\Omega_{\Lambda}=0.72$, $h\equiv
H_{0}/(100\,\mathrm{km\ s}^{-1}\mathrm{Mpc}^{-1})=0.701$,
$\sigma_{8}=0.812$, and a cosmic baryon fraction of $f_{b}=0.1658$,
following the updated values of the cosmological parameters from
\citet{komatsu09}. We also leave the values of the free parameters
associated with the galaxy formation model fixed to the fiducial
values given in Paper I. We check that these parameters still produce
good agreement with our calibration observations in the new ``hot
enrichment'' models in Section~\ref{sec:results:check}.

\section{Results}

In this section, we present our model results for the abundance ratios
and metallicities of the ICM, as well as some basic properties of
clusters, and compare them with a variety of observations. The
simulations were run on a grid of haloes with virial mass ranging from
$10^{14.25}\msun$ to $10^{15.6}\msun$ at an output redshift of
$z=0.05$.  This value was chosen because it is the mean redshift of
the groups and clusters in the observational samples.

\subsection{Cluster masses, temperatures and gas fractions}

\begin{figure}
\includegraphics[width=85mm]{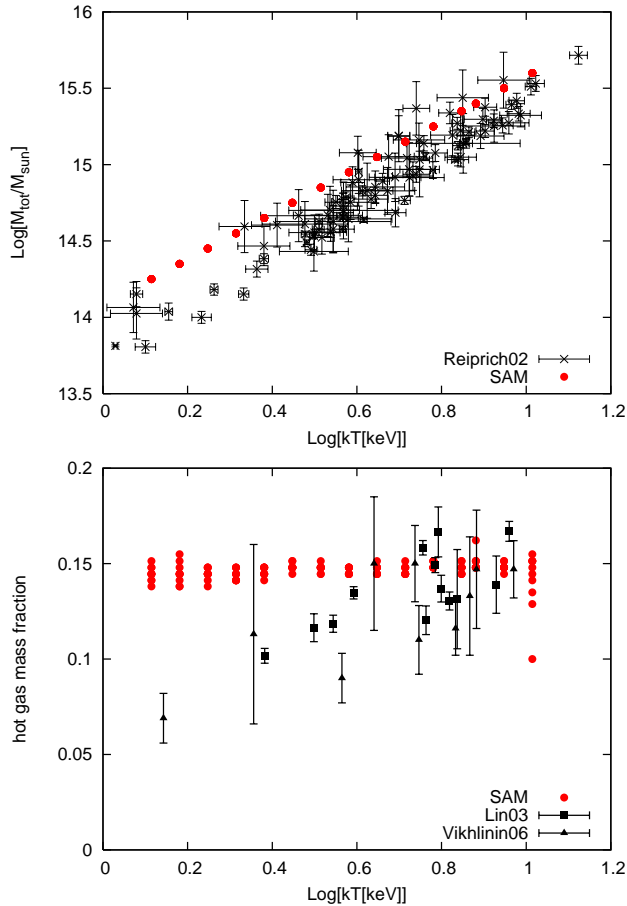}
\caption{{\it Top:} The virial mass--temperature relation for large
  groups and clusters. Red circles: Model; black crosses: data points
  from \citet{Reiprich02}. {\it Bottom:} The relation between the
  baryonic mass fraction and temperature. Red circles: Model; black
  triangles: data points from \citet{Vikhlinin06}; black squares: data
  points from \citet{Lin03}.}
\label{mass}
\end{figure}

In the observations, clusters are characterised by the measured
spectroscopic X-ray temperature. In the models, the ICM temperature is
taken to be equal to the halo virial temperature. Assuming
isothermality, this temperature relates to the virial velocity as
\begin{equation}
T_{\mathrm{vir}}[\mathrm{keV}]=35.9k_B(V_{\mathrm{vir}}[\mathrm{km/s}])^2,
\end{equation}
where $k_B$ is the Boltzmann constant. At redshift $z=0.05$, for the
chosen cosmology, the previous formula translates to $T_{\mathrm
  vir}=4.12(M_{\rm vir}/10^{15}\msun)^{2/3}\,\mathrm{keV}$.  The
virial temperature is, however, systematically lower than the X-ray
spectral temperature computed from the data \citep[by typically
  10\%,][]{bower08}. This small correction should not pose an issue in
the present work since the predicted and observed chemical properties
of the ICM show an extremely flat dependence on cluster temperature.

Before looking into the metal abundances and ratios, we study the
total mass and baryonic content of the simulated clusters. In Figure
\ref{mass} we show the virial mass and the baryonic gas fraction of
our simulated clusters as a function of temperature. The data points
are taken from \citet{Reiprich02} for the total mass and \citet{Lin03}
and \citet{Vikhlinin06} for the gas fraction. We do not show the
models with ``hot enrichment'' here since this affects only the metal
content and has a negligible effect on the total gas mass. In both
cases, the models are in qualitative agreement with the data, albeit
with some discrepancies. The small difference in the slope of the
mass-temperature relation arises because real clusters are not
strictly isothermal, as the models assume. Furthermore, correcting for
the 10\% systematic offset due to using the virial temperature rather
than the X-ray temperature would bring the models into better
agreement with the data. The hot gas fraction
($M_{\mathrm{hot}}/M_{\mathrm{vir}}$), however, shows no dependence
with temperature over this range, unlike the data, which shows a mild
trend of increasing baryonic fraction with temperature. This behavior
was already seen in S08 (their Figure 8). \citet{bower08} have shown
that if ``radio mode'' AGN feedback not only prevents the cooling of
gas but is also allowed to eject some of the hot gas out of the halo,
lower-mass clusters in the models will also show lower gas
fractions. It is worth noting that our models agree well with the {\it
  mean} gas fraction of the entire data sample and that model haloes
below 1 keV ($M_{\mathrm{vir}}\sim 10^{12}\msun$) show a sudden drop
in the predicted gas fraction (see S08 Figure 8). This step-like
behaviour in the gas fraction is common to other models
\citep{deLucia04,menci06}, and is due to the rapid transition from
infall-limited cooling (sometimes called ``cold mode'') to
cooling-time limited cooling (``hot mode'').

\subsection{Metallicities and abundance ratios}

\begin{figure}
\includegraphics[width=85mm]{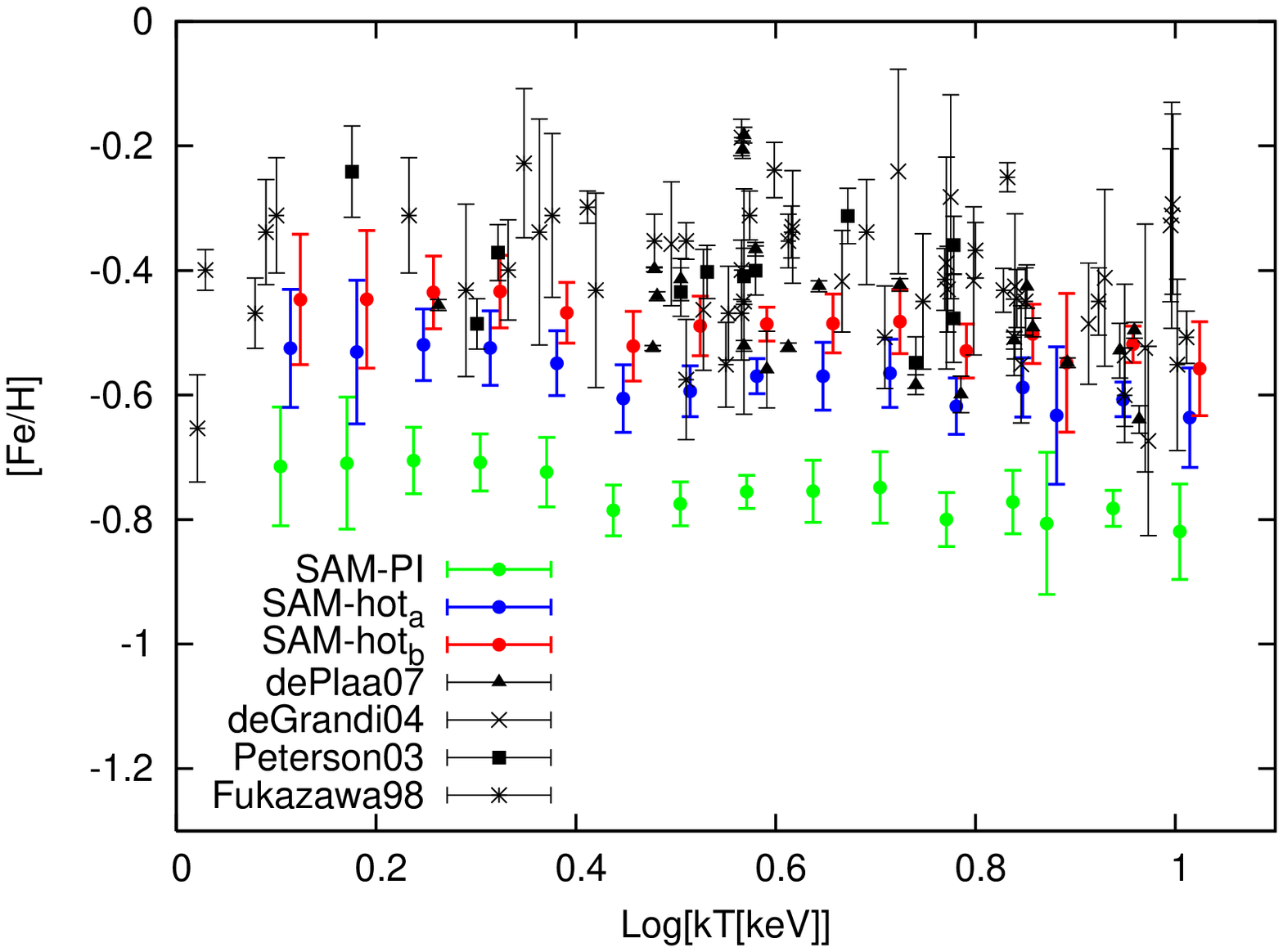}
\caption{Iron abundance in the ICM as function of temperature.  Green
  circles: Paper I fiducial model; blue and red circles: ``hot
  enrichment'' models with $A=0.03$ and $A=0.04$ respectively; black
  squares: data points from \citet{peterson03}; black triangles: data
  points from \citet{dePlaa07}; black stars: data points from
  \citet{fukazawa98}; black crosses: data points from
  \citet{degrandi04}. The errorbars on the observational data
  represent uncertainties, while for the models they indicate the mean
  and $1\sigma$ dispersion over different halo realizations.}
\label{iron}
\end{figure}

\begin{figure*}
\includegraphics[angle=270, width=170mm]{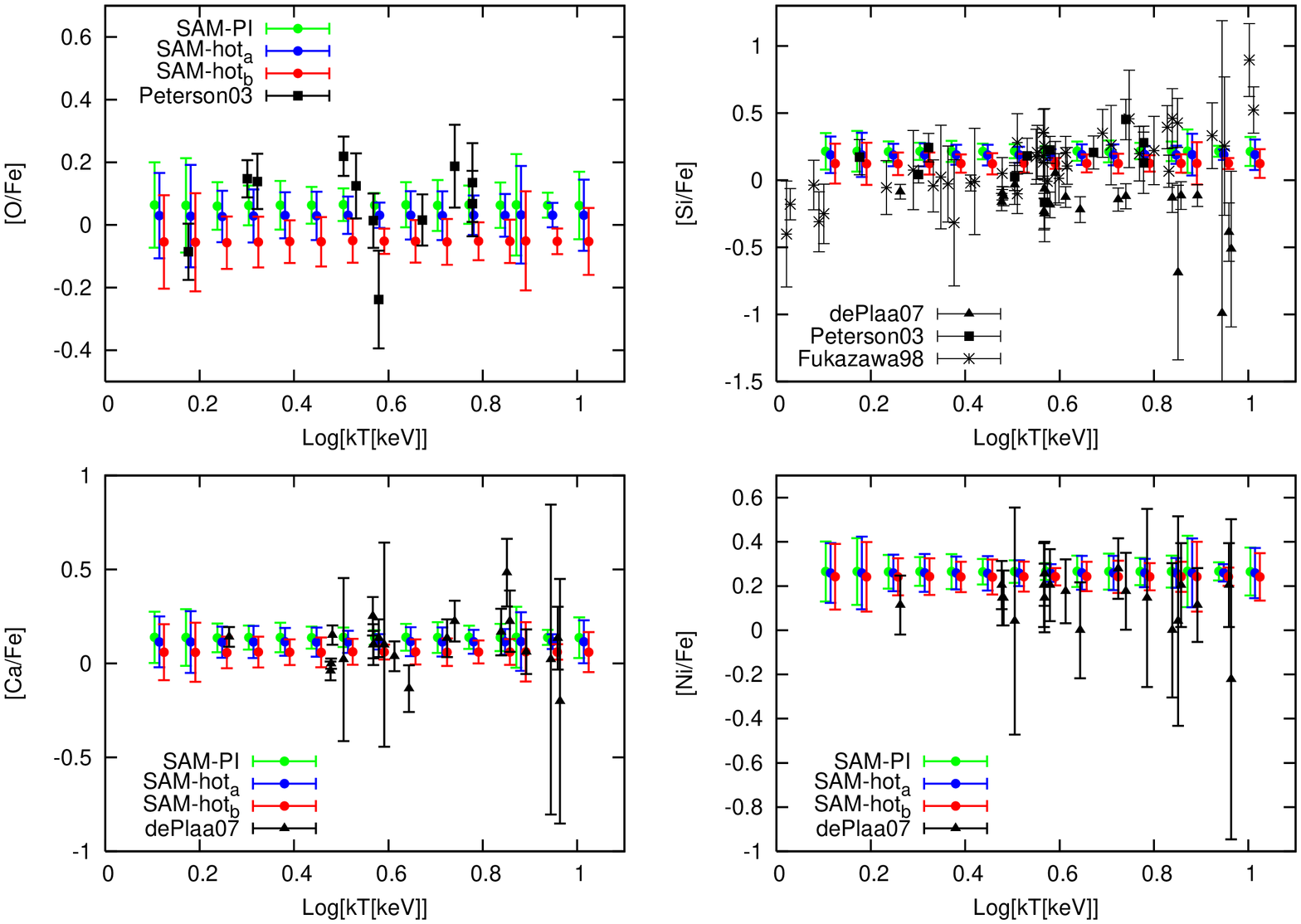}
\caption{Abundance ratios in the ICM as function of temperature, for
  ratios matched well by the models. Clockwise from top left: [O/Fe],
  [Si/Fe], [Ni/Fe] and [Ca/Fe]. Symbols as in Figure~\ref{iron}. The
  errorbars on the observational data represent uncertainties, while
  for the models they indicate the mean and $1\sigma$ dispersion over
  different halo realizations. }
\label{abu1}
\end{figure*}

\begin{figure*}
\includegraphics[angle=270, width=170mm]{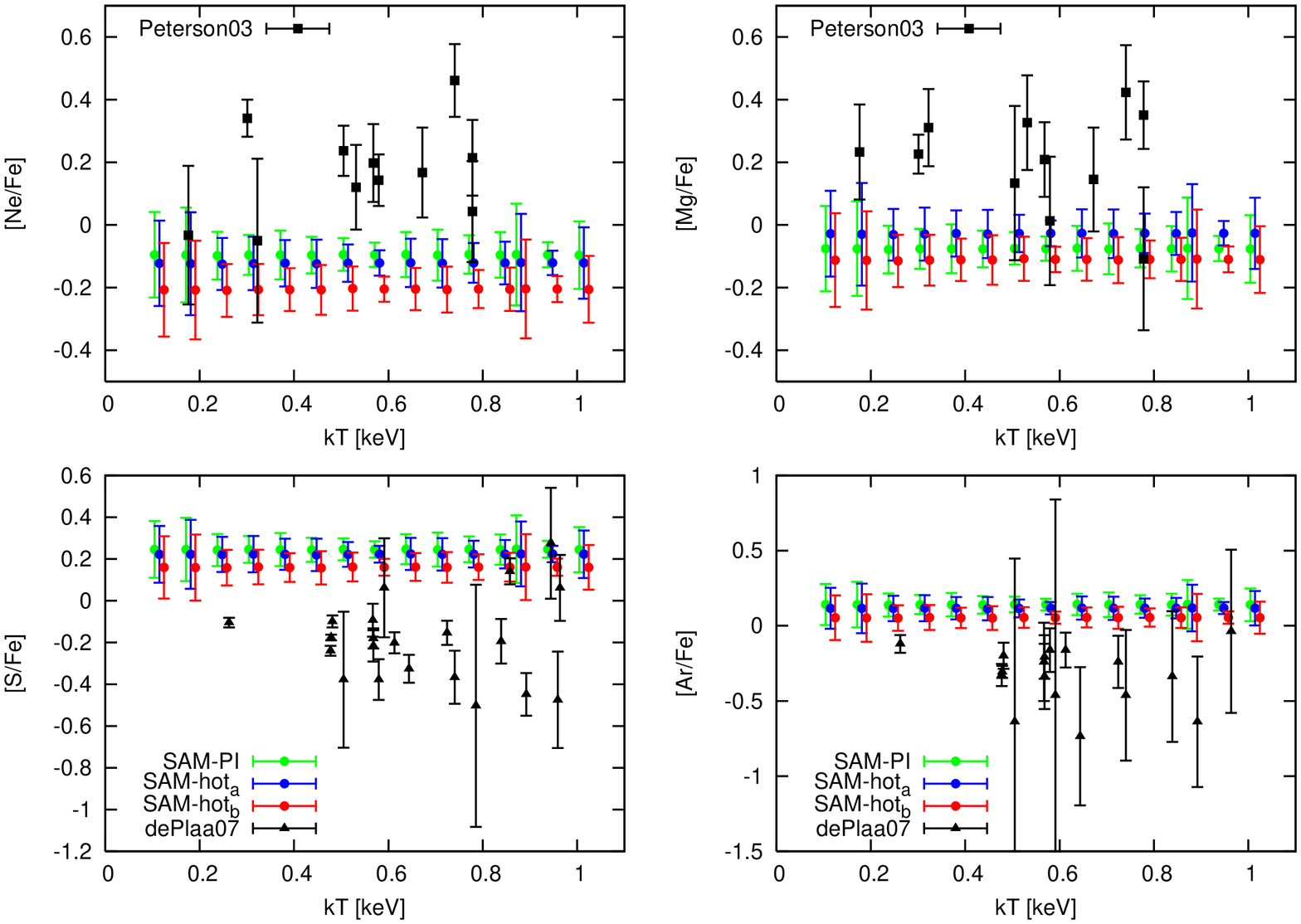}
\caption{Abundance ratios in the ICM as function of temperature, for
  ratios matched poorly by the models. Clockwise from top left:
  [Ne/Fe], [Mg/Fe], [Ar/Fe] and [S/Fe]. Symbols as in
  Figure~\ref{iron}. The errorbars on the observational data represent
  uncertainties, while for the models they indicate the mean and
  $1\sigma$ dispersion over different halo realizations. }
\label{abu2}
\end{figure*}

Our model proved in Paper I to be successful at reproducing the
metallicity- and [$\alpha$/Fe]-mass relations of local early-type
galaxies, as well as the SN rates as a function of SSFR.  We now
examine the iron content and the abundance ratios between different
elements and Fe in the ICM to further test its accuracy.  We use an
ensemble of X-ray cluster surveys for this purpose.  From
\citet{fukazawa98} we take Si and Fe; from \citet{peterson03} we take
O, Ne, Mg, Si and Fe; from \citet{degrandi04} we take Fe \citep[see
  also][]{ettori02}; and from \citet{dePlaa07} we take Si, S, Ar, Ca,
Fe and Ni. Galaxy clusters often show metallicity gradients for some
elements, with increasing abundances towards the cluster centre. These
clusters, known as {\it cool core} (CC) clusters, are mostly relaxed
systems and the central metal enhancement is associated with feedback
from the BCG. In contrast, {\it non-cool core} (NCC) clusters have
almost flat abundance profiles and show signatures of recent merging
events \citep{DGM01}. Since the observational data are measured near
the clusters centres and our models predict global abundances averaged
over an entire cluster, it is necessary to correct the observations
for gradients, but only for those clusters tagged as having {\it cool
  cores}.  We do this by following the procedure of \citet[Appendix
  A]{Nagashima05a}, who used the results of \citet{degrandi04} on Fe
gradients to convert the measured central Fe abundance to global
average values. Elements like Si, S, Ar, Ca and Ni are known to have
the same gradients as Fe, and are corrected by the same factor. On the
other hand, O, Ne and Mg do not show gradients even in CC clusters, so
we assume that the global abundance is equal to the central
measurement \citep{tamura01,tamura04}. We have also renormalised the
abundances to the solar values of \citet{Grevesse}, as in the models.

In Figure \ref{iron}, we examine the elemental abundance of iron
([Fe/H]). We pay particular attention to Fe because it is the ICM
element most precisely measured and most extensively studied. Both the
data and the models show a flat behaviour with temperature, an effect
also seen in the abundance ratios (see below).  It is clear that in
the original model, the iron abundance is too low and inconsistent
with the observations. Depositing the metals directly into the hot
halo gas (hot enrichment) appears to be a plausible solution, bringing
the models into marginal agreement with the data. In the models presented 
here, we have set the fraction of metals deposited directly into the ICM 
to 80\% ($f_{\mathrm{hot\, enrich}}=0.8$). The metallicity of the hot gas 
depends weakly on this parameter, incresing by about a factor 1.5 over the 
full parameter range (zero to one).Therefor e such a high value for 
$f_{\mathrm{hot\, enrich}}$ is necessary to have a significant effect. 
In this sense, also, further increasing its value beyond $\simeq 0.75$--0.8 
only raises the predicted abundances by a negligible amount. Another way to 
increase the iron content is by changing the number of type Ia supernovae by
adjusting the parameter $A$, which sets the fraction of binaries that
give rise to a SN Ia event. In Figure \ref{iron}, we show results for
both $A=0.03$, the fiducial value adopted in Paper I
($\mathrm{SAM-hot_a}$), and a slightly higher value, $A=0.04$
($\mathrm{SAM-hot_b}$). We see that indeed, a higher fraction of type
Ia SN binaries provides a better match to the ICM iron abundances,
however the value of this parameter is constrained by the observed SN
rates and can not take arbitrarily high or low values. As we show
later, a binary fraction of 0.04 is still consistent with SNe Ia rates
as a function of SSFR for galaxies in the local Universe while
producing ICM [Fe/H] abundances that are in marginal agreement with
the observed values.

In Figures \ref{abu1} and \ref{abu2} we show the abundance ratios of
different elements to iron (O, Ne, Mg, Si, S, Ar, Ca and Ni) in models
with and without ``hot enrichment''.  For those with ``hot recycling''
we again explore two different values for the SNIa binary fraction,
$A=0.03$ (${\rm SAM-hot_a}$) and $A=0.04$ (${\rm SAM-hot_b}$). There
are two aspects that are common to all the elements. Firstly, models
with hot enrichment show slightly lower abundance ratios than the
standard model, especially for $\alpha$ elements. This is not
surprising since the extra metals deposited by this mechanism come
mainly from very low mass galaxies that have low values of
[$\alpha$/Fe]. Secondly, all of the model abundance ratios show a flat
behaviour with temperature, as does the data, although the zero point
may disagree in some cases. Some of the ratios ([O/Fe], [Si/Fe],
[Ca/Fe] and [Ni/Fe]) show a very good match to the observations. On
the other hand, the predicted values of [Mg/Fe] and [Ne/Fe] are too
low, while [Ar/Fe] and [S/Fe] are overpredicted, although argon is
marginally consistent with the data.

The effect of the higher SNIa binary fraction is naturally stronger on
those elements produced mostly by SNII. With the higher [Fe/H]
required for a consistent iron abundance, the ratios [Ne/Fe] and
[Mg/Fe] are even lower and depart further from the observations.
[O/Fe] also decreases but is still consistent. [Ar/Fe] and [S/Fe] are
closer to the data but are still overpredicted and only marginally
consistent with the observations.  [Si/Fe] and [Ca/Fe] remain in good
agreement. Finally, [Ni/Fe] shows no variation with the binary
fraction parameter $A$, which is reassuring as both elements are
predominantly produced by SNIa.

\begin{figure}
\includegraphics[width=85mm]{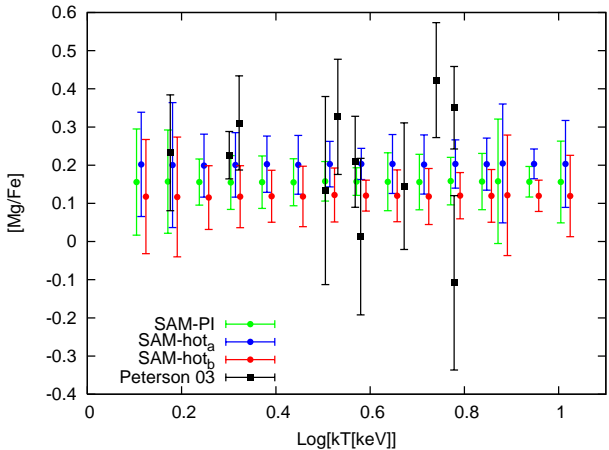}
\caption{[Mg/Fe] in the ICM as a function of temperature for models
  with the magnesium yield from SNII increased by a factor
  $2.5$. Symbols as in Figure~\ref{iron}. The errorbars on the
  observational data represent uncertainties, while for the models
  they indicate the mean and $1\sigma$ dispersion over different halo
  realizations. }
\label{2xMg}
\end{figure}

In the case of [Mg/Fe], the model ratios can be raised by increasing
the Mg yield in stars above 20 \msun, a common practice with the WW95
yields \citep[see, e.g., ][]{Francois04}. In Paper I there was no need
for such a modification, but in this case increasing the Mg yield by a
factor of 2.5 brings the ICM abundance ratio into good agreement with
the data, while still maintaining an observationally consistent
[Mg/Fe] in the galaxies' stellar component. A slightly higher factor
would give a better match for the ICM, but in that case the stellar
abundance ratios would be too high. Models with a boosted magnesium
yield are shown in Figure \ref{2xMg}. In principle, the same exercise
could be done with the yields of other elements that are
underpredicted (Ne) or overpredicted (Si, Ar). However this should not
be considered a solution, but simply a tentative constraint on
nucleosynthesis from the chemical evolution models. Also, we have
assumed that the different elements in the ICM either have no radial
gradients at all or that they have the same gradient as iron (for which
there are fairly good measurements). This simple assumption might not
be strictly true and a more accurate correction for gradients could
bring the models and the data into better agreement. Future
observations of gradients of elements other than iron in the ICM would
shed some light on this matter.

\subsection{Effects of ``hot enrichment'' on galaxy properties}
\label{sec:results:check}

\begin{figure*}
\includegraphics[angle=270, width=170mm]{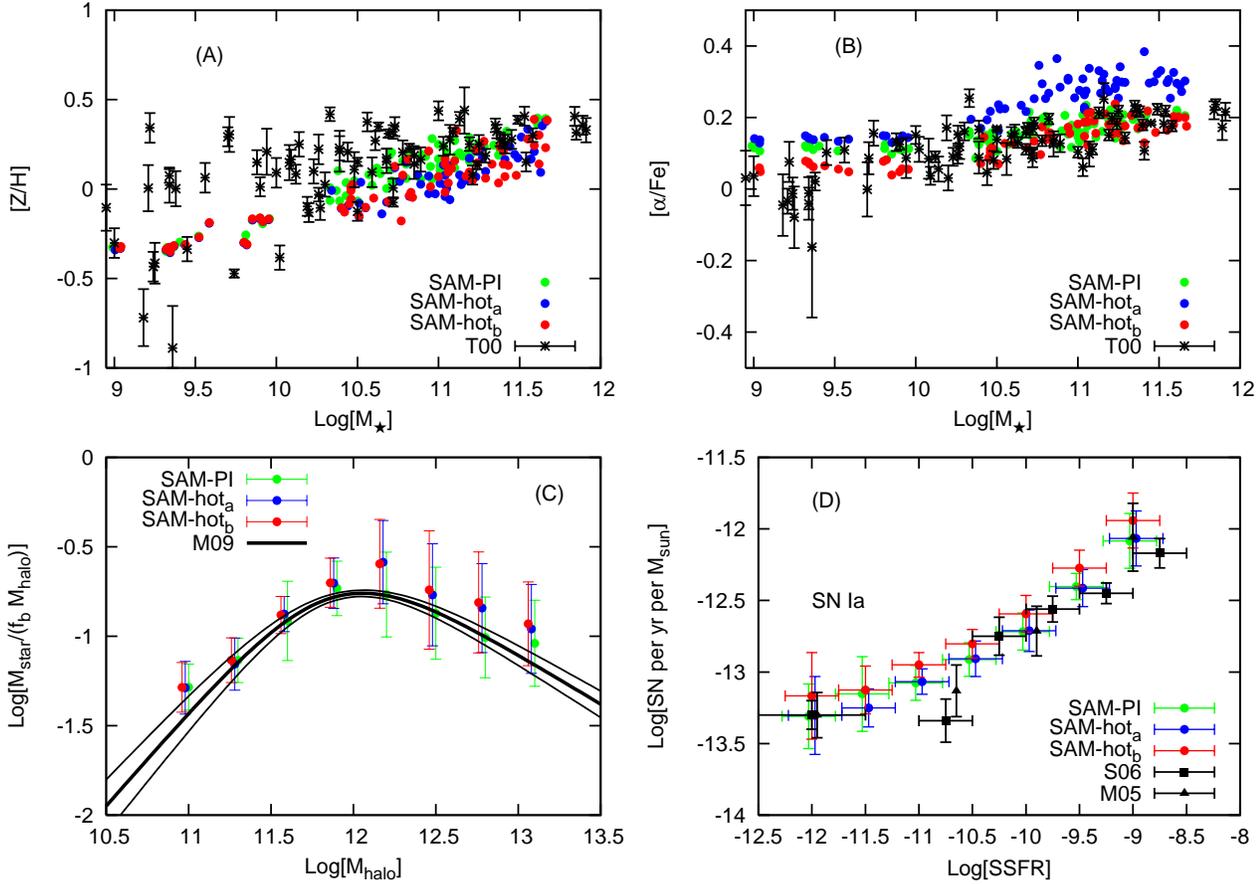}
\caption{Predicted properties of galaxies in the local
  Universe. Clockwise, starting from the top left panel: (A) [Z/H]
  vs. stellar mass for early-type galaxies; (B) [$\alpha$/Fe]
  vs. stellar mass for early-type galaxies; (C) Fraction of baryons in
  the form of stars as a function of halo mass; (D) Type Ia SNR
  vs. SSFR.  Symbols -- green circles: Paper I fiducial model; blue
  and red circles: ``hot enrichment'' models with $A=0.03$ and
  $A=0.04$ respectively; T00: reanalysed metallicities and abundance
  ratios from \citet{T00a} as presented in Paper I; M05 and S06: SN
  rates from \citet{Mannucci05} and \citet{sullivan06} respectively;
  M09: the empirical relation, and $1\sigma$ uncertainties, derived by
  \citet{Moster08}.}
\label{gxs}
\end{figure*}

We have introduced a fairly significant modification to our model ---
the deposition of the majority of the newly produced metals into the
hot halo gas, instead of into the cold interstellar gas. It is
important to check whether this change has an impact on the properties
of galaxies that we used to calibrate our previous models. In Figure
\ref{gxs} we show the same three models as in the previous section
(Paper I fiducial, SAM-PI; hot enrichment with $A=0.03$,
$\mathrm{SAM-hot_a}$; and hot recycling with $A=0.04$,
$\mathrm{SAM-hot_b}$). We show the metallicity, [$\alpha$/Fe] ratio,
and SN Ia rate of galaxies, and compare them with the same data
samples from the local Universe as in Paper I. We remind the reader
that the fiducial model from Paper I had $A=0.03$.
 
The metallicities of early-type galaxies are not significantly
affected by this change and remain in agreement with the observations
(panel A).  However, galaxies in models with ``hot enrichment'' have
their [$\alpha$/Fe] increased (especially the most massive galaxies).
A higher value of the binary fraction parameter $A$ brings the
abundance ratios back into agreement with the observations (panel
B). From panel (D) in Figure \ref{gxs}, we see that the lower value of
$A=0.03$ provides a better fit to the SN Ia rates as a function of the
specific star formation rate (SSFR), although a value of $A=0.04$ is
still consistent with the data. Considering that a higher fraction of
stars that explode as SNe Ia is also required to make the iron content
in the ICM consistent with the data, $A=0.04$ gives a better overall
agreement between the models and the observations.

Another concern is that when metals are deposited directly into the
hot gas, elevating the metal content, the cooling rate increases,
possibly resulting in the conversion of a larger fraction of the baryons
in the halo into stars. This could result in the production of an
overabundance of massive galaxies relative to observations. We check this by
investigating the ratio of the mass of baryons that have turned into
stars in the central galaxy to the mass of baryons that would be
contained in the halo in the absence of star formation or feedback
(i.e. $f_{\mathrm{b}} M_{\mathrm{vir}}$, where $f_{\mathrm{b}}$ is the
universal baryon fraction), as a function of halo mass. We compare the
model predictions with the empirical constraint from \citet{Moster08},
which is derived by requiring that the observed stellar mass function
is reproduced for a given assumed multiplicity function of dark matter
halos (i.e. as predicted by a given $\Lambda$CDM model). As we can see
from panel (C) in Figure \ref{gxs}, the effect of the new model
ingredients is small and all models are consistent with each other and
with the data.

\section{Discussion and Conclusions}

We have investigated the metal enrichment of the intracluster medium 
within the framework of hierarchical assembly using the same model 
presented in Paper I, which successfully reproduces the abundance ratios 
of early-type galaxies in the local Universe by assuming a slightly 
flat IMF ($x=1.15$) and a bimodal Delay-Time-Distribution of type Ia 
supernovae.

Our most important finding is the need for some form of metal enriched
outflows from galaxies because the ICM iron abundance is too low
otherwise. Adopting ``hot enrichment'', in which 80\% of the
metal-rich material ejected by the stars is deposited directly into
the ICM rather than the ISM, seems to be a reasonable solution.  We
also need slightly more type Ia supernovae, both for the iron in the
ICM and the [$\alpha$/Fe] in the galaxies. Although the fit to SNR
vs. SSFR is not as good as in Paper I, it is still consistent with the
observations.

Regarding the elemental abundance ratios in the ICM, the models
predict flat behaviour with cluster temperature, in agreement with the
observations. For some elements (O, Si, Ca, Ni) the zero-point is
reproduced remarkably well, while others agree only marginally (Ar,
S), or are significantly underpredicted (Ne, Mg). This occurs
irrespective of whether ``hot enrichment'' is assumed or not. The
[Mg/Fe] can be fixed by increasing the Mg yield in SN II (as is
commonly done with the WW95 yields). The discrepancy in the other
elements may arise from uncertainties in the yields and/or the
correction for radial gradients (we assume that elements that have a
gradient share the same one as Fe, which might not be strictly
correct, although they cannot be too different).

Overall the model {\it simultaneously} produces acceptable predictions
for the chemical properties of galaxies in the local Universe and the
ICM in nearby clusters. This is yet another step forward in building a
self-consistent framework for predicting the properties of diverse
populations within the context of the hierarchical galaxy formation
framework.

\section*{Acknowledgements}

We thank the directors of the Max-Planck-Institut f\"ur Astronomie,
H.-W.~Rix, and the Kapteyn Astronomical Institute, J.M.~van der Hulst,
and NOVA, the Lorentz Center and the Leids Kerkhoven-Bosscha Fonds for
providing travel support and working space during the gestation of
this paper. We also thank the Space Telescope Science Institute for
travel funding and hospitality. We thank Y.S. Li and B.K. Gibson
for helpful discussions.

\bibliographystyle{mn2e}
\bibliography{models}

\appendix

\label{lastpage}

\clearpage
\end{document}